\PassOptionsToPackage{unicode}{hyperref}
\PassOptionsToPackage{hyphens}{url}
\PassOptionsToPackage{dvipsnames,svgnames,x11names}{xcolor}
\documentclass[
  11pt,
]{article}
\usepackage{amsmath,amssymb}
\usepackage{iftex}
\ifPDFTeX
  \usepackage[T1]{fontenc}
  \usepackage[utf8]{inputenc}
  \usepackage{textcomp} 
\else 
  \usepackage{unicode-math} 
  \defaultfontfeatures{Scale=MatchLowercase}
  \defaultfontfeatures[\rmfamily]{Ligatures=TeX,Scale=1}
\fi
\usepackage{lmodern}
\ifPDFTeX\else
\fi
\IfFileExists{upquote.sty}{\usepackage{upquote}}{}
\IfFileExists{microtype.sty}{
  \usepackage[]{microtype}
  \UseMicrotypeSet[protrusion]{basicmath} 
}{}
\makeatletter
\@ifundefined{KOMAClassName}{
  \IfFileExists{parskip.sty}{%
    \usepackage{parskip}
  }{
    \setlength{\parindent}{0pt}
    \setlength{\parskip}{6pt plus 2pt minus 1pt}}
}{
  \KOMAoptions{parskip=half}}
\makeatother
\usepackage{xcolor}
\providecommand{\tightlist}{%
  \setlength{\itemsep}{0pt}\setlength{\parskip}{0pt}}
\usepackage[margin=1in]{geometry}
\usepackage[most]{tcolorbox}
\usepackage{xurl}
\ifLuaTeX
  \usepackage{selnolig}  
\fi
\IfFileExists{bookmark.sty}{\usepackage{bookmark}}{\usepackage{hyperref}}
\IfFileExists{xurl.sty}{\usepackage{xurl}}{} 
\hypersetup{
  pdftitle={Data Center Governance Under AI Expansion Pressures},
  pdfauthor={Yacine Jernite, Sasha Luccioni, Bruna Treveling, Angela Lee, Carolina Molina, Xinying Chu},
  colorlinks=true,
  linkcolor={black},
  filecolor={Maroon},
  citecolor={Blue},
  urlcolor={blue!60!black},
  pdfcreator={LaTeX via pandoc}}

\title{Data Center Governance Under AI Expansion Pressures}
\usepackage{etoolbox}
\makeatletter
\providecommand{\subtitle}[1]{
  \apptocmd{\@title}{\par {\large #1 \par}}{}{}
}
\makeatother
\subtitle{Technical Report}
\author{Yacine Jernite\\
  \small\href{mailto:yacine@huggingface.co}{yacine@huggingface.co}
\and Sasha Luccioni\\
  \small\href{mailto:sasha@sustainableaigroup.com}{sasha@sustainableaigroup.com}
\and Bruna Trevelin\\
  \small\href{mailto:bruna@huggingface.co}{bruna@huggingface.co}
\and Angela Lee\\
  \small\href{mailto:angela.lee@nyu.edu}{angela.lee@nyu.edu}
\and Carolina Molina\\
  \small\href{mailto:cm7368@nyu.edu}{cm7368@nyu.edu}
\and Xinying Chu\\
  \small\href{mailto:xc3405@nyu.edu}{xc3405@nyu.edu}}
\date{}

\begin{document}
\maketitle

{
\hypersetup{linkcolor=}
\setcounter{tocdepth}{3}
\tableofcontents
}
\clearpage
\hypertarget{executive-summary}{%
\section*{Executive Summary}\label{executive-summary}}
\addcontentsline{toc}{section}{Executive Summary}

\begin{tcolorbox}[breakable,colback=gray!6,colframe=gray!50,boxrule=0.5pt,arc=2pt]

The promised benefits of data centers are global, but their negative impacts are very much local, affecting the energy bills, aquifers and living environments of citizens across the United States. While governments and authorities have historically provided data centers with incentives such as tax credits and exemptions, AI's aggressive growth in scale has changed this dynamic, and recent months have seen them issue
bills and file lawsuits in order to ensure that data centers remain a net positive influence on their legislatures. Existing pieces of federal legislation -- such as the Public Trust Doctrine, the Pollution Prevention, Clean Air and Safe Drinking Water Acts -- exist to manage natural resources, but their scope remains contested, and the current federal government's stance on competitiveness and growth of AI has put the onus on regional governments to endeavor to protect local resources. There are also state- and regional laws and regulations that are relevant for building and operating data centers since, in practice, they require substantial land, electricity, water, and infrastructure at a local level. Given this situation, communities are currently provided
with the opportunity to raise their concerns regarding the potential negative impacts of how data centers are built and operated, and have existing regulations that can be used to put pressure on data center operators.

\textbf{Key Insights:}

\begin{itemize}
\tightlist
\item
  \textbf{Federal Priorities vs.~Local Autonomy:} A paradigm shift is occurring in American industrial policy, highlighted by a July 2025 Executive Order designed to fast-track AI data centers, which   weakens cornerstone environmental protections to eliminate bureaucratic bottlenecks. This top-down governance model prioritizes computational power and national security over resource
  conservation, effectively neutralizing the legal levers communities use to govern heavy development.
\item
  \textbf{The Soaring Burden on Ratepayers:} Data center electricity consumption is skyrocketing, with projections indicating these   facilities could consume up to 17\% of U.S. electricity by 2030.   Because these massive facilities require new power plants and delivery   infrastructure, the financial burden can be shifted onto local residential ratepayers. For example, recent approvals in Georgia allowed for the construction of new methane gas units primarily to power data centers, raising concerns about cost-shifting to residents.
\item
  \textbf{Severe Environmental Toll:} Beyond raw energy consumption, data centers heavily strain local resources. Large data centers can consume between 5 to 9 million gallons of water per day for cooling,
  the equivalent of a small town's usage. Furthermore, communities  living near these facilities face increased noise pollution, as well as air pollution from the installation of methane gas turbines and  backup diesel generators.
\item
  \textbf{A Growing Legislative Backlash:} In response to these   unchecked pressures, states and municipalities are aggressively   mobilizing to regulate the industry. In 2025 and early 2026, hundreds
  of state-level bills were introduced to demand transparency, mandate  water and energy reporting, and adjust tax exemptions. In states like  New York, South Dakota, and Oregon, legislators have even proposed  moratoriums on new hyperscale data center construction to study their impact on utility rates and local resources.
\item
  \textbf{The Push for Transparency:} Communities are demanding that tech companies and developers act as ``good neighbors'' by sharing  project details in plain language and ensuring operators pay the full  costs of infrastructure.  Consequently, community groups are
  increasingly leveraging lawsuits across multiple states to force   developers and local governments to comply with open records requests   and rigorous environmental reviews.
\end{itemize}

\end{tcolorbox}

\hypertarget{i.-introduction}{%
\section{I. Introduction}\label{i.-introduction}}

\hypertarget{a.-motivation-and-goals}{%
\subsection{A. Motivation and Goals}\label{a.-motivation-and-goals}}

As data centers are increasingly being built across the U.S. to
facilitate artificial intelligence (AI) development and deployment, the costs to local communities
have become apparent. State governments and local authorities
incentivize the development of data centers through tax exemptions,
credits, and re-zoning, enticed by promises of job creation and
investment. However, before the jobs and investments
materialize,\footnote{\href{https://www.foodandwaterwatch.org/wp-content/uploads/2026/01/RB_2601_DataCenterJobs.pdf}{Artificial
  Jobs: The Illusion of Big Tech's Data Center Employment Claims}} local
residents must bear the burdens of higher energy costs, environmental
pollution, amongst other issues such as potential gaps in revenue,
urgent, large-scale construction and short-term labor. While state and
local governments previously sought to attract technology companies to
build data centers to revitalize areas, the scale and urgency of AI
means that communities may now find themselves finally having the
opportunity to negotiate, rather than needing to attract.\footnote{\href{https://www.brookings.edu/articles/turning-the-data-center-boom-into-long-term-local-prosperity/}{Turning
  the data center boom into long-term, local prosperity \textbar{} Brookings Institute}} This also means that communities have the opportunity to voice their concerns about how data centers are built and operate in their neighborhoods -- lawsuits and
regulations could indicate a shift in the dynamic.

Data center development comes with serious implications for the local
residents and authorities of the localities where they are built.
Communities see an increase in utility rates, high water and energy
consumption, straining resources and environmental
challenges.\footnote{\href{https://stpp.fordschool.umich.edu/sites/stpp/files/2025-07/stpp-data-centers-2025.pdf}{What Happens When Data Centers Come to Town?}} The current federal
government has emphasized its interest in remaining highly competitive
in this industry and has left it to states and municipalities to address
the concerns mentioned above. States and local legislators, however, are
increasingly motivated by having data centers ``pay their fair
share.''\footnote{\href{https://www.politico.com/newsletters/power-switch/2026/03/13/data-center-tax-breaks-split-virginia-democrats-00828380}{Data   center tax breaks split Virginia Democrats - POLITICO}} Thus, the
issuance of bills and filing of lawsuits by state organizations and even
residents directly affected by the development have been the strongest
efforts to address the environmental, resource and financial problems.

\hypertarget{1.-public-trust-and-industrial-influence}{%
\subsubsection{1. Public Trust and Industrial
Influence}\label{1.-public-trust-and-industrial-influence}}

Environmental concerns always raise issues related to public goods and
the public interest. It is important to recognize that there are
different parties at play in the development of data centers: these
include developers, technology companies, federal, state and local governments
and the residents of the communities where they are located. The Public
managing certain natural resources for the benefit of the public; more
recently, this doctrine has helped frame arguments around whether state
governments should take more aggressive actions against climate change.
In one of the most recognized cases where this doctrine has been
applied, which is \emph{Juliana v. U.S.}, No.~18-36082 (9th Cir.
2020), the plaintiffs argued that the federal government had violated
the doctrine by encouraging and allowing the use of fossil fuels. However, the
federal government has consistently argued in response that the PTD is
limited to state regulation, and that the climate or atmosphere would
fall outside of its scope.\footnote{\href{https://nationalaglawcenter.org/the-public-domain-basics-of-the-public-trust-doctrine/}{The
  Public Domain: Basics of the Public Trust Doctrine -- National
  Agricultural Law Center}}

Technology is constantly evolving: the world has seen many industries surge and transform their products into essential
elements for our economy, development and even for our daily lives. One
definition given to an essential industry is that it ``is considered
necessary for the nation's economy and that may be protected or
supported by the government.''\footnote{\href{https://dictionary.cambridge.org/us/dictionary/english/essential-industry}{ESSENTIAL
  INDUSTRY definition \textbar{} Cambridge English Dictionary}} For instance, the telecommunications sector is considered integral for personal and
commercial interactions; it continues to bring with it multiple environmental issues related to energy consumption, e-waste and resource depletion, carbon emissions, water use, etc. Despite those negative impacts, state governments identify it as essential infrastructure; in the state of New York, executive order 202.6 as a response to Covid-19 identified telecommunications as essential infrastructure and therefore manufacturing and data centers as essential infrastructure as
well.\footnote{\href{https://esd.ny.gov/guidance-executive-order-2026}{Guidance
  on Executive Order 202.6 \textbar{} Empire State Development}} While Artificial Intelligence is increasingly discussed as a core strategic imperative by the federal government in the US, it is unclear whether it is yet a necessary component of an essential industry or an essential
industry in its own right.

\hypertarget{2.-federal-laws-and-regulations}{%
\subsubsection{2. Federal Laws and
Regulations}\label{2.-federal-laws-and-regulations}}

There are a number of federal regulations that aim to protect the environment from invasive practices and exploitation of natural resources. The Federal Land Policy and Management Act of 1976 declares
for public lands to be managed in a way that protects the quality of scientific, scenic, historical, environmental, air, water resource and archaeological values, recognizing at the same time the importance of public  involvement.\footnote{\href{https://www.blm.gov/sites/default/files/AboutUs_LawsandRegs_FLPMA.pdf}{The  Federal Land Policy and Management Act of 1976 as amended}} The
Pollution Prevention Act, on the other hand, focuses its attention on reducing the amount
of pollution through cost-effective changes in production, operation and raw materials use; it also includes practices that increase efficiency in the use of energy, water or other natural resources, in an effort to protect them through conservation.\footnote{\href{https://www.epa.gov/laws-regulations/summary-pollution-prevention-act}{Summary
  of the Pollution Prevention Act \textbar{} US EPA}}

Key statutes such as the Clean Air Act require facilities to comply with
air quality standards and obtain permits for emissions, which are directly relevant to data center backup generation and energy
use\footnote{\href{https://www.epa.gov/stationary-sources-air-pollution/clean-air-act-resources-data-centers}{Clean Air Act Resources for Data Centers \textbar{} US EPA}}. The Clean
Water Act similarly requires permits for pollutant discharges and
construction-related impacts on water bodies, including stormwater
permits and Section 404 permits for land development activities
affecting wetlands.\footnote{\href{https://www.epa.gov/laws-regulations/summary-clean-water-act}{Summary
  of the Clean Water Act \textbar{} US EPA}}\footnote{\href{https://www.epa.gov/cwa-404/permit-program-under-cwa-section-404}{Permit
  Program under CWA Section 404 \textbar{} US EPA}} In addition, the
National Environmental Policy Act (NEPA) requires environmental review
for projects involving federal funding, approvals, or land, making it a
key procedural step in large-scale infrastructure siting\footnote{\href{https://www.epa.gov/nepa/epa-compliance-national-environmental-policy-act}{EPA
  Compliance with the National Environmental Policy Act \textbar{} US
  EPA}}. Together, these statutes form a permitting framework that developers must navigate before construction can proceed. In practice, compliance with air, water, and environmental review requirements
functions as a gatekeeping mechanism, determining whether projects can move forward and under what conditions.

In addition to these core statutes, other federal laws also play a supporting role in regulating environmental and resource impacts
associated with data center development. For instance, the Safe Drinking Water Act
(SDWA) establishes standards for the quality of public drinking water and may be implicated where large-scale water withdrawals or potential contamination risks are involved. Similarly, federal policies such as the Taxpayer Bill of Rights reflect broader principles of fairness and accountability in government decision-aking, which can inform debates over cost allocation and whether infrastructure and energy burdens should be borne by private developers or shifted onto the public.

These permitting standards are not applied mechanically. Federal agencies retain discretion in determining whether projects satisfy regulatory requirements, particularly in assessing environmental
impacts, infrastructure costs, and system reliability -- a principle
long recognized by courts in cases such as \emph{Chevron U.S.A., Inc.~v.
NRDC}, 467 U.S. 837 (1984) and \emph{Baltimore G. \& E. Co.~v. NRDC}, 462
U.S. 87 (1983). As a result, approval decisions inherently involve judgment, and projects may be conditioned, delayed, or effectively denied where their impacts are considered unacceptable.

Recent policy developments reflect the evolving role of federal
regulation in this space. For instance, legislative proposals such as the Clean Cloud
Act\footnote{\href{https://www.congress.gov/bill/119th-congress/senate-bill/1475}{S.1475
  - Clean Cloud Act of 2025 119th Congress (2025-2026)}} and executive
actions promoting AI infrastructure\footnote{\href{https://www.whitehouse.gov/presidential-actions/2025/01/removing-barriers-to-american-leadership-in-artificial-intelligence/}{Removing
  Barriers to American Leadership in Artificial Intelligence -- The
  White House}} suggest a growing tension between environmental constraints and economic priorities. Taken together, federal regulation shapes data center development indirectly by setting standards that must be interpreted and applied in context, embedding discretion within the  approval process rather than relying on explicit mechanisms of refusal.

\hypertarget{3.-state-and-local-laws-and-regulations}{%
\subsubsection{3. State and Local Laws and
Regulations}\label{3.-state-and-local-laws-and-regulations}}

At the state and local level, environmental regulation of data center development often operates through land use controls, permitting processes, and environmental review (see the American Planning Association's primer on data center siting and local regulation, and EPA's overview of NEPA).\footnote{\href{https://www.planning.org/planning/2021/summer/data-centers-evolved-a-primer-for-planners/}{Data Centers Evolved: A Primer for Planners \textbar{} APA}} Although data centers are often described as digital infrastructure, they are also large physical facilities that require substantial land, electricity,
water, and local infrastructure (see the American Planning Association's
discussion of data center planning considerations).\footnote{\href{https://www.planning.org/planning/2021/summer/data-centers-evolved-a-primer-for-planners/}{Data
  Centers Evolved: A Primer for Planners \textbar{} APA}} For that reason, local governments can shape their development through zoning and land use rules that determine where data centers may be built and under what conditions (see the American Planning Association's primer on
zoning and permitting for data centers).

Permitting processes provide a second layer of control. Beyond the question of where a data center may be located, projects may need
site-plan, building, air, water, or other approvals before construction can proceed (see EPA's pages on Clean Air Act permitting and NPDES permitting).\footnote{\href{https://www.epa.gov/caa-permitting}{Permitting Under the Clean Air Act \textbar{} US EPA}} These processes matter because they govern how a project is built in practice and may allow
agencies to impose project-specific conditions, while also creating some
opportunities for public review and comment in jurisdictions where environmental review applies (see EPA's explanation of the NEPA review process)~\footnote{\href{https://www.epa.gov/nepa/national-environmental-policy-act-review-process}{National
  Environmental Policy Act Review Process \textbar{} US EPA}}. However,
these procedures do not always give communities strong power to block a
project, and they often function more as mechanisms for delay, modification, or conditional approval than as clear paths to outright refusal (this is an inference from the structure of zoning, permitting, and environmental review described in these sources).\footnote{\href{https://www.planning.org/planning/2021/summer/data-centers-evolved-a-primer-for-planners/}{Data
  Centers Evolved: A Primer for Planners \textbar{} APA}}

Environmental review and environmental compliance add a further layer, but serve a distinct function from permitting: rather than directly granting or denying approval, they require agencies to assess and disclose the broader environmental impacts of development before key decisions are made (see EPA's overview of NEPA and NEPA review
process).\footnote{\href{https://www.epa.gov/nepa/what-national-environmental-policy-act}{What is the National Environmental Policy Act? \textbar{} US EPA}} Although data centers may appear less visibly polluting than traditional industrial facilities, they can still create significant environmental
burdens through high electricity and water use, generator emissions, and site-specific impacts on stormwater and surrounding infrastructure (see the American Planning Association's primer, EPA's Clean Air Act
permitting page, and EPA's NPDES materials).\footnote{\href{https://www.planning.org/planning/2021/summer/data-centers-evolved-a-primer-for-planners/}{Data
  Centers Evolved: A Primer for Planners \textbar{} APA}} Environmental
review therefore helps make these costs visible and quantifiable in formal decision-making and\footnote{\href{https://www.epa.gov/nepa/national-environmental-policy-act-review-process}{National
  Environmental Policy Act Review Process \textbar{} US EPA}} may provide an important channel through which local residents and advocacy groups can press for mitigation measures (see EPA's NEPA materials, including its citizen-oriented guidance). At the same time, environmental review is not always a strong barrier to development -- it often operates more as a tool of disclosure and mitigation than as a
direct basis for stopping a project (again, this is an inference from how EPA describes the review process rather than an express statement in a single source). In terms of financial and tax regulations, states have also adopted regulations that aim to protect citizens, but also, companies. Refer to the Appendix~\ref{appendix} for an analysis of example state law regulations directly related to environmental and financial tools and regulation.

\hypertarget{4.-executive-orders-and-the-ratepayer-pledge}{%
\subsubsection{4. Executive Orders and the Ratepayer
Pledge}\label{4.-executive-orders-and-the-ratepayer-pledge}}

On July 23, 2025, the Executive Order titled ``Accelerating Federal
Permitting of Data Center Infrastructure'' was published. Its purpose is
to usher in a ``golden age'' of American technological dominance by
aggressively fast-tracking the construction of AI data centers and the
immense energy infrastructure required to power them, with the
administration declaring this rapid buildout a national security and
economic priority. The EO specifically targets ``Qualifying Projects'',
which are defined as data centers or related component manufacturing
facilities that require over 100 megawatts (MW) of new power, involve
over \$500 million in capital expenditures, or are deemed critical to
national security. To support these projects, the order broadens the
definition of necessary infrastructure to include ``dispatchable
baseload energy sources,'' explicitly listing natural gas, coal, and
nuclear power, alongside heavy electrical grids.

To achieve this rapid expansion, the EO deploys three main strategies:

\begin{enumerate}
\def\labelenumi{\arabic{enumi}.}
\tightlist
\item
  \textbf{Financial Incentives:} It directs the Department of Commerce
  and other agencies to provide financial support for these projects,
  including loans, grants, and tax incentives.\\
\item
  \textbf{Federal Land Utilization:} It mandates the Department of the
  Interior, the Department of Energy, and the Department of Defense to
  open up federally-owned lands and military installations to lease them
  to data center developers.\\
\item
  \textbf{Regulatory Relief:} It revokes a prior AI infrastructure
  order (EO 14141) and orders sweeping rollbacks of environmental review
  processes across multiple federal agencies to eliminate bureaucratic
  bottlenecks.
\end{enumerate}

This changes the data center governance landscape in several ways --
first of all, previously, data center governance was largely a localized
issue, managed by municipal zoning boards, regional utility grid
operators, and state environmental regulators balancing community
resources. This EO fundamentally shifts governance to a top-down,
federalized model that treats AI infrastructure as a matter of national
security. By heavily incentivizing and mandating the buildout of coal,
gas, and nuclear baseloads specifically to serve AI, the order
acknowledges that the existing grid cannot handle the AI boom.
Consequently, it strips away local and environmental friction, pushing a
``build-first'' governance model that prioritizes raw computational
power and geopolitical technological supremacy over resource
conservation.

The order also dismantles the environmental review processes that have
governed US industrial development for the last half-century. It
effectively bypasses or severely weakens several cornerstone
environmental protection laws, including:

\begin{itemize}
\tightlist
\item
  \textbf{The National Environmental Policy Act (NEPA):} The EO bypasses
  the previously required comprehensive Environmental Impact Statements
  by mandating agencies to widely adopt ``categorical exclusions''
  (exemptions usually reserved for minor projects with no environmental
  impact). It also creates a loophole by declaring that any project
  receiving less than 50\% of its funding from the federal government is
  no longer considered a ``major Federal action,'' effectively shielding
  multi-billion-dollar, privately funded data centers from NEPA review
  entirely.\\
\item
  \textbf{The Clean Air Act (CAA) \& Clean Water Act (CWA):} The EO
  directs the EPA to alter existing CAA and CWA regulations to
  explicitly expedite AI projects. Furthermore, it pushes the Army Corps
  of Engineers to issue ``nationwide permits'' for water use, replacing
  rigorous, site-specific water impact reviews with high-level
  approvals.\\
\item
  \textbf{The Endangered Species Act (ESA):} The EO scraps localized
  biological consultations of endangered species in favor of 10-year
  ``programmatic'' consultations, which are generic reviews that allow
  developers to bypass specific wildlife impact studies.\\
\item
  \textbf{CERCLA (Superfund) and Brownfields:} Rather than waiting for
  exhaustive environmental remediation, the EO fast-tracks the
  repurposing of highly contaminated Superfund and Brownfield sites for
  data center use, prioritizing rapid industrial deployment over
  prolonged environmental cleanup.
\end{itemize}

Overall, this EO represents a paradigm shift in American industrial
policy, with the current administration moving to sacrifice decades-old
environmental protections in exchange for a hypothetical dominance in
the so-called ``AI race''. By neutralizing laws like NEPA and ESA, the
order removes the primary legal levers communities and conservationists
have historically used to govern, delay, or mitigate heavy industrial
development, paving the way for the unrestrained expansion of AI
infrastructure.

\hypertarget{ii.-efforts-and-implementation}{%
\section{II. Efforts and
Implementation}\label{ii.-efforts-and-implementation}}

\hypertarget{a.-transparency:-importance-and-reality.}{%
\subsection{A. Transparency: Importance and
Reality}\label{a.-transparency:-importance-and-reality.}}

Since data centers are seen as necessary infrastructure to support digital
life, their responsible development should include clear and ongoing dialogue with the communities they are impacting. Considering their strains on water,
energy, land and financial resources, it is important for them to approach local authorities and residents as ``good neighbors'' -- this means listening to concerns, sharing project details in plain language, thoughtful site design, landscaping, setbacks, transparent reporting on energy and water use and commitments for the use of community resources, as well as commitments to keep utility costs low for ratepayers and ensure operators pay the full costs of electricity and
infrastructure.\footnote{\href{https://www.uschamber.com/technology/data-centers-deliver-local-value}{Data  Centers Deliver Local Value \textbar{} US Chamber of Commerce}}

Making dialogue a primary component of data center development is important;
however, in many cases, efforts to that end have not met to the expectation of residents, or have been altogethr missing. For instance, in Minnesota, the suits filed by Minnesota Center for Environmental
Advocacy (MCEA) make this discomfort obvious: in a case filed against City of Pine Island in the District Court of the State of Minnesota, the MCEA challenged the city's environmental review of a new  development, which would take up to 3 million square feet of land and up to 2,700 megawatts of electrical capacity for a Google data center~\footnote{\href{https://www.mncenter.org/legal-updates-mceas-data-center-cases}{Legal updates on MCEA's data center cases}}. The court ruled in the MCEA's favor in May 2026 and in October of the same year, the project was halted until 
a supplemental environmental assessment would be conducted through the Minnesota Pollution Control Agencym which would delay construction by at least two years~\footnote{\href{https://www.mncenter.org/ensuring-data-center-development-follows-minnesota-law}{Ensuring Data Center Development Follows Minnesota Law}}.

A similar dynamic is playing out in Wisconsin and Virginia -- Midwest Environmental
Advocates filed a lawsuit in the Wisconsin Circuit Court for Dane County
against the Wisconsin Public Service Commission, claiming that the
Commission unlawfully denied an open records request for unredacted
documents on the projected electrical load of Meta's Beaver Dam data
center, which is one of three data center projects responsible for a projected 40 percent increase in statewide energy demand.\footnote{\href{https://www.wpr.org/news/lawsuit-expected-energy-demand-meta-data-center-beaver-dam}{Lawsuit
  seeks release of expected energy demand from Meta data center in
  Beaver Dam}} In August 2026, a judge ruled that the Commission had acted unlawfully in denying the request and ordered the information released.\footnote{\href{https://wislawjournal.com/2026/09/11/judge-wisconsin-psc-unlawfully-withheld-data-center-records/}{Judge:
  PSC unlawfully withheld data center records \textbar{} Wisconsin Law
  Journal}}. In Virginia, the Prince William County Circuit Court invalidated rezoning approvals for a data center development due to the county's failure to comply with Virginia Code, including late advertisement of the hearings and failure to make the rezoning documents public and proceeding despite legal
warnings.\footnote{\href{https://www.beankinney.com/data-center-infrastructure-under-siege-lessons-from-virginias-digital-gateway-decision}{Data
  Center Infrastructure Under Siege: Lessons from Virginia's Digital
  Gateway Decision \textbar{} Bean, Kinney \& Korman}}; in March 2026, a panel of judges at the Court of Appeals of Virginia confirmed the ruling.\footnote{\href{https://virginiamercury.com/2026/04/08/va-court-of-appeals-stops-major-data-center-development-in-prince-william-county/}{Va.
  Court of Appeals stops major data center development in Prince William
  County \textbar{} Virginia Mercury}} A month later, the Board of County Supervisors voted to stop defending the rezoning, after spending roughly \$1.7 million in public funds on the litigation, and abandoned its appeal.\footnote{\href{https://papers.ssrn.com/sol3/papers.cfm?abstract_id=7418899}{Case Study: Prince William County — Governance Capture, the Revolving Door, and the Democratic Correction}}.

These lawsuits have begun to produce concrete results, although largely
through procedural leverage rather than rulings on the merits. For instance, the Hermantown data center litigation illustrates both the value and the limits of
such arrangements -- in MCEA v.\ City of Hermantown, the MCEA agreed
in March 2026 to stay its lawsuit while the city updated its environmental review, and the developer agreed not to begin construction without notifying MCEA in time for it to seek an injunction.\footnote{MCRO\_69DU-CV-25-3448}\footnote{\href{https://www.mncenter.org/ensuring-data-center-development-follows-minnesota-law}{Ensuring
  Data Center Development Follows Minnesota Law \textbar{} Minnesota
  Center for Environmental Advocacy}} After the city adopted a second environmental assessment and approved a special permit for the project in October 2026, MCEA and the resident group Stop the Hermantown Data Center filed a new suit, alleging that the updated review still fails to adequately assess the project's impacts on air quality, noise, electronic waste, water and greenhouse gas emissions.\footnote{\href{https://www.startribune.com/hermantown-data-center-google-lawsuit/601910810}{Opponents
  sue Hermantown again over Google data center \textbar{} Star Tribune};
  \href{https://bringmethenews.com/minnesota-news/hermantown-data-center-after-councils-go-ahead-opponents-appeal-over-environmental-review}{Hermantown
  data center: After council's go-ahead, opponents appeal over
  environmental review \textbar{} Bring Me The News}}. 
  
\hypertarget{b.-ongoing-state-level-regulatory-efforts}{%
\subsection{B. Ongoing State-Level Regulatory
Efforts}\label{b.-ongoing-state-level-regulatory-efforts}}

In 2025, there were significant developments in data center related bills at a state level. A total of 267 bills were considered, of these, 126 addressed energy use, considering topics such as negotiating special rates, nuclear energy and studies to understand the impact of data
centers.\footnote{\href{https://s3.amazonaws.com/multistate.us/production/resources/r5icxD60PGfum2Xbt/attachment/State\%20Data\%20Center\%20Policy\%20101\%20--\%20MultiState\%20Resource\%20\%2812.4.25\%29.pdf}{State Data Center Policy 101 Legislative Trends, Key Terms, and What to Expect in 2026}} In 2026, more than 300 state legislation bills had been filed as of
February -- in fact, State governments are currently the quickest growing source of regulations. They provide tax incentives for data center development, while also imposing regulations to ensure that residential utility customers do not end up bearing the burden of increased energy use.\footnote{See generally: 
  \href{https://www.gibsondunn.com/when-data-center-developers-have-options-state-regulatory-treatment-is-key-to-success/}{When
  Data Center Developers Have Options, State Regulatory Treatment Is Key
  to Success - Gibson Dunn}} However, environmental reporting requirements vary
by state:

\begin{enumerate}
\def\labelenumi{\arabic{enumi}.}
\tightlist
\item
  \textbf{Virginia}
\end{enumerate}

In Virginia, the country's largest data center market, numerous bills
have been introduced to tighten regulation, especially in reference to
transparency.

\textbf{HB 284} (to amend Code of Virginia by adding section 56-596.7 to
chapter 23 of title 56): directs Dominion Energy (DE) and Appalachian
Power (AP) to file a petition with the State Corporation Commission for
approval of demand flexibility programs that apply to high energy demand
customers; it also requires DE and AP to file status reports on their
demand flexibility programs with the State Corporation Commission 3
years after initial program approval and every 3 years thereafter, and
requires the Commission to submit information summarizing the status and
performance of such programs.\footnote{\href{https://lis.virginia.gov/bill-details/20261/HB284}{HB284
  - 2026 Regular Session \textbar{} LIS}}

\textbf{HB 591} (to amend Code of Virginia by adding section
45.2-1706.2) aims to establish a policy of the commonwealth that seeks
to i) promote coordination between data centers, state agencies and
regional grid operators to share information on energy usage,
interconnection timelines and barriers to rapid deployment of renewable
and flexible energy resources; ii) incentivize data centers to
participate in demand response programs; iii) encourage flexible energy
practices; iv) ensure large scale data centers pay their fair share of
infrastructure investments, mitigating impact on residents; v) require
data centers to report water and energy usage as well as sustainability
measures and vi) promote cybersecurity, physical security and supply
chain security.\footnote{\href{https://lis.virginia.gov/bill-details/20261/HB591}{HB591
  - 2026 Regular Session}}

In relation to this, \textbf{SB43} (to amend Code of Virginia by adding
section 51.5-39.14 to chapter 8.1 of title 51.5) directs the department
of energy to evaluate and assess benefits and impacts of best practices
and recommendations for demand response programs in the commonwealth, as
well as to submit a report of such evaluation by November 1,
2026.\footnote{\href{https://lis.virginia.gov/bill-details/20261/SB43}{SB43
  - 2026 Regular Session \textbar{} LIS}}

Locally, communities are seeing that data centers bring enough drawbacks
to necessitate new limits and conditions. Legislators in Virginia,
including in Loudoun and Fairfax County, have proposed zoning and
tax bills that seek to regulate data centers.\footnote{\href{https://www.gibsondunn.com/when-data-center-developers-have-options-state-regulatory-treatment-is-key-to-success/}{When
  Data Center Developers Have Options, State Regulatory Treatment Is Key
  to Success - Gibson Dunn}} Loudoun County, home to one of the largest
concentrations of data centers in the world, is enacting zoning
standards for new data centers and utility substations.\footnote{\href{https://www.loudoun.gov/5990/Data-Center-Standards-Locations}{Data
  Center Standards \& Locations \textbar{} Loudoun County, VA - Official
  Website}} Fairfax County, a prosperous county close to D.C., has also
enacted zoning ordinances about the location, size, design, and noise
levels for new data centers.\footnote{\href{https://www.fairfaxcounty.gov/news/board-supervisors-approve-new-data-center-zoning-ordinance-amendment}{Board
  of Supervisors Approves New Data Center Zoning Ordinance Amendment}}

In March 2026, state legislators also brought forward bills that sought
to address the impact of data centers. \textbf{SB 553} (to amend and
reenact section 62.1-44.38 of the Code of Virginia) addresses water use,
directing water utilities to report at least quarterly how much water is
going to data centers.\footnote{\href{https://lis.virginia.gov/bill-details/20261/SB553}{SB553
  - 2026 Regular Session \textbar{} LIS}} \textbf{HB 507} (to amend the
Code of Virginia by adding section 10.1-1322.6 to Article 1 of chapter
13 of title 10.1) mandates that the state's Department of Environmental
Quality deny air permits for data center generators unless they pass
certain air pollution and technology thresholds.\footnote{\url{https://lis.virginia.gov/bill-details/20261/HB507}
  \url{https://lis.virginia.gov/bill-details/20261/HB507/text/HB507}}
\textbf{SB 253} (to amend and reenact section 56-585.1:2 of the Code of
Virginia) includes a provision that allows the State Corporation
Commission to approve shifting more energy costs onto facilities, such
as data centers, instead of onto residential consumers.\footnote{\href{https://lis.virginia.gov/bill-details/20261/SB253}{SB253
  - 2026 Regular Session \textbar{} LIS}} Debates over the sales and use
tax exemption for data centers contributed to lawmakers failing to
finalize a budget, which will only be finalized with a special session
in late April 2026. While the state Senate wants to remove the
exemption, arguing it causes a revenue loss of \$1.6 billion annually,
the House wants to keep the exemption (but with heightened environmental
and energy requirements to qualify).\footnote{\href{https://virginiamercury.com/2026/03/17/data-center-bills-dominated-this-years-general-assembly-heres-what-passed/}{Data
  center bills dominated this year's General Assembly. Here's what
  passed | Virginia Mercury}}

At the same time, former Governor Youngkin's State of the Commonwealth
speech in January 2025 emphasized a desire to keep attracting data
center development, touting economic opportunities for local
communities.\footnote{Id.} Virginia currently continues to offer retail
sales and use tax exemptions for use in data centers.\footnote{P. 191
  \href{https://www.doa.virginia.gov/reports/ACFReport/2025/2025-ACFReport-for-web-Entire-Report.pdf}{Annual
  Comprehensive Financial Report for the Fiscal Year Ended June 30,
  2025}} The state's Department of Environmental Quality has guidelines
specifically granting permits to air pollution emissions from data
centers' primary and backup power generators.\footnote{\url{https://www.deq.virginia.gov/home/showpublisheddocument/27430/638730583843800000}}

\begin{enumerate}
\def\labelenumi{\arabic{enumi}.}
\setcounter{enumi}{1}
\tightlist
\item
  \textbf{Other states: New York, South Dakota, Oregon.}
\end{enumerate}

New York proposes \textbf{Bill A10141} (to amend the environmental
conservation law by adding 31 §§31-0101 - 31-0107) and \textbf{S9144}
(to amend the environmental conservation law by adding 31 §§31-0101 -
31-0107) which imposes a moratorium on the issuance of permits for new
data centers and requires the public service commission to issue an
order to minimize their impact on electricity and gas rates for
residential, commercial and industrial users. These bills are currently
in committee assembly and in committee senate respectively.\footnote{\href{https://www.nysenate.gov/legislation/bills/2025/A10141}{NY
  State Assembly Bill 2025-A10141};
  \href{https://www.nysenate.gov/legislation/bills/2025/A10141}{NY State
  Assembly Bill 2025-A10141};
  \href{https://www.nysenate.gov/legislation/bills/2025/S9144}{NY State
  Senate Bill 2025-S9144};
  \href{https://www.nysenate.gov/legislation/bills/2025/S9144}{NY State
  Senate Bill 2025-S9144}}

South Dakota has introduced \textbf{Senate Bill 232,} which imposes a
one-year moratorium on the construction or expansion of hyperscale data
centers.\footnote{\href{https://sdlegislature.gov/Session/Bill/27269/301228}{2026
  Senate Bill 232 \textbar{} South Dakota Legislature}} Similarly,
Oregon's \textbf{senate bill 1488} proposes a three-year moratorium on
data centers with an electrical load greater than 100MW, to allow for
the study of the center's impact on water supply, utility rates and
property values.\footnote{\href{https://www.multistate.us/insider/2026/2/20/state-data-center-legislation-in-2026-tackles-energy-and-tax-issues}{State
  Data Center Legislation in 2026 Tackles Energy and Tax Issues
  \textbar{} MultiState}}

\hypertarget{iii.-cost-and-environmental-outcomes-in-data-center-permitting}{%
\section{III. Cost and Environmental Outcomes in Data Center
Permitting}\label{iii.-cost-and-environmental-outcomes-in-data-center-permitting}}

\hypertarget{a.-managing-energy-cost-shifts}{%
\subsection{A. Managing Energy Cost
Shifts}\label{a.-managing-energy-cost-shifts}}

A 2024 Department of Energy study projected that data center electricity
consumption would rise from 176 terawatt-hours (TWh) in 2023 to between
325 and 580 TWh by 2028.\footnote{\href{https://www.congress.gov/crs-product/R48762}{Data
  Center Energy Infrastructure: Federal Permit Requirements \textbar{}
  Congress.gov}} While data centers currently use around 5\% of U.S.
electricity, research by the Electric Power Research Institute shows
that by 2030, driven by AI usage, data centers could use as much as 17\%
of U.S. electricity.\footnote{\href{https://www.nytimes.com/2026/03/04/climate/data-centers-electricity-trump.html}{Why
  Tech Firms Are Working With Trump on Data Centers and Energy Costs}}
The increased use of electricity means that costs of serving large,
energy-intensive customers such as hyperscale data centers are
potentially partially borne by local, residential ratepayers. Because
these data centers require so much energy, new power plants and other grid infrastructure must be built, and if the bills these data centers pay do not cover the cost,
then residential ratepayers must cover the difference.\footnote{\href{https://www.nytimes.com/2026/03/04/climate/data-centers-electricity-trump.html}{Why
  Tech Firms Are Working With Trump on Data Centers and Energy Costs}}

Currently, there are no legally binding federal energy standards that
apply to data centers in the private sector.\footnote{\href{https://www.congress.gov/crs-product/R48646\#fn57}{Data
  Centers and Their Energy Consumption: Frequently Asked Questions
  \textbar{} Congress.gov}} Rather, states essentially grant monopolies
to investor-owned utilities, which are for-profit companies such as AEP,
ConEd, Southern Company (Georgia Power), to deliver energy to
ratepayers. Government-approved electricity prices reimburse utilities
for their operational expenses and provide utilities an opportunity to
earn a fixed rate of return on their capital investments. Since utility
companies are insulated from many ordinary business risks and shielded
from competitive pressures, public utility regulators, or PUCs, must
protect the public from a utility's monopoly power and motivate the
company to provide reliable and cost effective service.\footnote{\href{https://eelp.law.harvard.edu/wp-content/uploads/2025/03/Harvard-ELI-Extracting-Profits-from-the-Public.pdf}{Extracting
  Profits from the Public: How Utility Ratepayers Are Paying for Big
  Tech's Power}} A Harvard study by the Environmental \& Energy Law
Program highlights the role utility companies play in increased energy
costs for residential ratepayers: utility companies are allowed
confidential treatment of contracts with data centers, utility companies
can find the gap between federal (FERC) and state regulations on cost
allocation schemes to assign more costs to residential ratepayers, and
data centers can pursue direct contracts with non-utility power
generators to bypass the utility-owned transmission network.\footnote{\href{https://eelp.law.harvard.edu/wp-content/uploads/2025/03/Harvard-ELI-Extracting-Profits-from-the-Public.pdf}{Extracting
  Profits from the Public: How Utility Ratepayers Are Paying for Big
  Tech's Power}}

As an example, the Georgia Public Service Commission (GPSC) approved a
new rule in January 2025 that allows the state-regulated utility,
Georgia Power, to charge data centers for electric service in a way that
protects the state's retail ratepayers from cost-shifting.\footnote{\href{https://psc.ga.gov/site/assets/files/8617/media_advisory_data_centers_rule_1-23-2025.pdf}{PSC
  Approves Rule to Allow New Power Usage Terms for Data Centers}} While
GPSC ensures that Georgians ``will not subsidize data centers,'' it also
allowed Georgia Power to build five new methane gas units totaling 9,985
MW of new energy generation, of which 80\% will be used to power data
centers.\footnote{\href{https://psc.ga.gov/site/downloads/datacenterfactsheet.pdf}{Data Center Fact Sheet}} Research by the Southern Environmental Law Center,
one of the groups filing a motion to reconsider, shows that since 2023,
the average residential Georgia Power bill has increased by more than
\$43 per month, while Georgia Power profits have increased.\footnote{\href{https://www.selc.org/press-release/psc-unanimously-votes-to-approve-georgia-powers-data-center-plan-without-sufficient-customer-protections/}{PSC
  unanimously votes to approve Georgia Power's data center plan without
  sufficient customer protections}} Groups against the Georgia Power
plan estimate that the total cost to ratepayers would be \$50 to 60
billion; during the hearing, the plan's cost structure remained unclear
and details apparently left for future cases to work out.\footnote{\href{https://cleanenergy.org/news/lame-duck-commission-sides-with-georgia-power-and-data-centers-over-families-and-small-businesses/}{Lame-duck
  Commission Sides with Georgia Power and Data Centers over Families and
  Small Businesses - Southern Alliance for Clean Energy (SACE)}}

In July 2025, the Public Utilities Commission of Ohio (PUCO) approved a
tariff by American Electric Power Ohio (AEP), the state's public utility
company, on data centers and ``mobile data centers'' (for mining
crypto). This updated tariff (Schedule DCT) subjects applicable data
centers to specified load ramp periods, longer contract terms, adjusted
minimum demand charges, collateral requirements, capacity reassignment
limitations, and a new service enrollment process, amongst other
provisions.\footnote{\url{https://dis.puc.state.oh.us/ViewImage.aspx?CMID=A1001001A25G09B43531I00509}}
However, the Ohio Manufacturers' Association Energy Group appealed the
PUCO decision in February 2026 based on the rationale that the tariff is
facially discriminatory and lacks evidentiary support, since they
believe the tariff is based on overstated speculative load forecasts by
AEP.\footnote{\href{https://www.supremecourt.ohio.gov/pdf_viewer/pdf_viewer.aspx?pdf=997940.pdf\&subdirectory=2025-1458\%5CDocketItems\&source=DL_Clerk}{AEP
  Data Center Tariff SCT Appellant Merit Brief (2025-1458)
  (945919-9).DOCX}}

Locally, towns are worried about energy prices and lacking the
infrastructure to support these data centers. Nelsonville City Council
in Ohio enacted Ordinance 84-25 because the energy use of data centers
was considered to be ``detrimental'' to the residents, who were
concerned about the strain on local resources.\footnote{\href{https://www.athensmessenger.com/news/data-centers-a-growing-source-of-controversy-in-ohio/article_bc880e93-fad5-4e47-b1e5-b577c0668d08.html}{Data
  centers a growing source of controversy in Ohio \textbar{} News
  \textbar{} athensmessenger.com}} Localities from Pennsylvania to
Virginia are bringing forth proposals to restrict data centers, with
places such as Loudoun County and York County in Virginia establishing
stricter processes for data centers to be
approved.\footnote{\href{https://www.govtech.com/policy/data-center-growth-driving-locals-to-push-for-more-say}{Data
  Center Growth Driving Locals to Push for More Say};}\footnote{\href{https://virginiamercury.com/2025/06/20/virginia-doesnt-have-statewide-data-center-regulations-localities-are-making-their-own-rules-instead/}{Virginia
  doesn't have statewide data center regulations. Localities are making
  their own rules.}}

\hypertarget{b.-environmental-protection}{%
\subsection{B. Environmental
Protection}\label{b.-environmental-protection}}

Data centers, often built in rural areas with abundant land, also have
significant environmental consequences. The mere operation of data
centers can cause noise, air, and water pollution.  Research shows that large data centers each
use 5 million gallons of water per day for cooling, equivalent to
a town of 10,000 to 50,000 people,\footnote{\href{https://www.eesi.org/articles/view/data-centers-and-water-consumption}{Data
  Centers and Water Consumption \textbar{} Article \textbar{} EESI}} and
even as much as 9 million as AI usage increases.\footnote{\href{https://www.nytimes.com/2025/07/14/technology/meta-data-center-water.html}{Meta
  Built a Data Center Next Door. The Neighbors' Water Taps Went Dry. -
  The New York Times}} A 2021 report estimated that data centers in the
U.S. consume 449 million gallons of water per day for cooling-- 163.7 billion
gallons annually.\footnote{\href{https://www.eesi.org/articles/view/data-centers-and-water-consumption}{Data
  Centers and Water Consumption \textbar{} Article \textbar{} EESI}}
 The need for more energy resources inevitably requires building new power plants and generators, which emit more
fossil fuels in the area. Environmental consequences have also led to health problems amongst residents near data centers, from noise-related hypertension and anxiety to nitrogen oxide (NOx) smog and formaldehyde-related respiratory
illnesses.\footnote{\href{https://thereader.mitpress.mit.edu/the-staggering-ecological-impacts-of-computation-and-the-cloud/}{The
  Staggering Ecological Impacts of Computation and the Cloud \textbar{}
  The MIT Press Reader}}

Noise in data center facilities can come from: cooling towers, air handling units, air-cooled chillers, indoor servers, and back-up diesel generators.\footnote{\href{https://ketchumandwalton.com/5-sources-of-noise-in-data-centers-how-to-control-it/}{5
  Sources of Noise in Data Centers \& How to Control it \textbar{}
  Ketchum \& Walton Co.}} Regulations at the federal level are fragmented because data centers are not yet categorized separately from heavy industry. Occupational Safety
and Health Administration (OSHA) Standard 1910.95 details the necessities of providing noise protection for sustained noise levels
above 85  decibels.\footnote{\href{https://www.osha.gov/laws-regs/regulations/standardnumber/1910/1910.95}{1910.95
  - Occupational noise exposure. \textbar{} Occupational Safety and
  Health Administration}} Though noise from data center facilities can be disruptive to neighbors, it has not surpassed the OSHA requirements;
they fluctuate in the range between 80 to 110 dB, not high enough for federal regulatory actions.\footnote{\href{https://ketchumandwalton.com/5-sources-of-noise-in-data-centers-how-to-control-it/}{Sources of Noise in Data Centers \& How to Control it \textbar{}
  Ketchum \& Walton Co.}}

Environmental groups have so far relied on the Clean Air Act and the
Clean Water Act in their lawsuits. NAACP and Southern Environmental Law
Center (SELC) sent a letter of intent to sue xAI over their unpermitted
installation and operation of methane gas turbines to power xAI's first
data center in South Memphis in 2025. xAI complied and removed some
turbines and acquired a permit for the remaining 15.\footnote{\href{https://www.selc.org/press-release/groups-threaten-lawsuit-over-xais-unpermitted-gas-turbines-in-mississippi/}{Groups  threaten lawsuit over xAI's unpermitted gas turbines in Mississippi -
  Southern Environmental Law Center}} NAACP and SELC then submitted an appeal in July 2025 to Memphis and Shelby County Air Pollution Control Board over the granting of a permit to xAI's data center to operate 15 methane gas turbines 24/7. Their argument states that
Shelby County erroneously believed that the Clean Air Act exempted xAI's
gas turbines.\footnote{\href{https://naacp.org/articles/naacp-and-advocacy-groups-appeal-permit-xais-south-memphis-data-center-decisions-around}{NAACP,
  Advocacy Groups Appeal Permit for xAI's South Memphis Data Center}} In
February 2026, SELC, Earthjustice, and NAACP sent another notice of
intent to sue xAI over Clean Air Act violations at its second data
center in South Memphis, but with its 27 turbines across state lines in
Mississippi.\footnote{\href{https://earthjustice.org/wp-content/uploads/2026/02/2026.02.13-final-xai-southaven-noi-with-exhibit-a.pdf}{Notice
  of Intent to Sue for Violations of the Clean Air Act at the Gas
  Turbines Power Plant Located at}}

State and local level actions are more effective at attempting to
curtail environmental harm, since noise pollution can be reduced via local
zoning ordinances and reporting requirements. In New Castle County,
Delaware, the local council enacted Ordinance No.~25-101 to regulate
data center zoning and water usage, requiring closed-loop systems for
cooling, which reuse water as much as possible.\footnote{\href{https://newcastlecode.portal.civicclerk.com/event/2420/files/attachment/11723}{New
  Castle County Council Agenda Center • CivicClerk}} Lawsuits at these
levels hinge on transparency -- alleging that cities and developers fall
short of reporting standards, for example in Minnesota Center for
Environmental Advocacy (MCEA)'s active state lawsuits.\footnote{\href{https://www.mncenter.org/latest-our-data-center-lawsuits-mceas-chief-legal-officer}{The
  latest on our data center lawsuits from MCEA's Chief Legal Officer
  \textbar{} Minnesota Center for Environmental Advocacy}}

\section{IV. Conclusion}\label{iv.-conclusion}

The rapid expansion of AI data centers has exposed a widening gap between where their global benefits accrue and where their local costs are borne. With the current federal government prioritizing speed of deployment and narrowing the reach of statutes such as NEPA and the Clean Air Act, the task of governing these facilities from an environmental standpoint has fallen largely to states, municipalities and the communities
that host them. This report has shown that existing legal tools, from the Public Trust Doctrine to state environmental policy acts and open records laws, still offer meaningful levers, but that their effectiveness depends heavily on transparency and on the ability of residents and advocacy groups to use them.

The litigation reviewed in Section II illustrates both the promise and the
limits of this approach -- procedural claims, rather than rulings on the substance of environmental harm, have produced the most concrete outcomes to date. However,  these victories remain project-specific, and settlements do not create binding precedent, leaving each community to fight the same battles anew. Legislative responses are beginning to fill this gap: state bills addressing energy cost allocation, water and energy reporting, demand flexibility and tax incentives, together with local zoning ordinances and proposed moratoriums, point toward a model in which data center operators bear the
full cost of the infrastructure they require and disclose the information communities need to evaluate them. Whether these efforts succeed will depend on states adopting clear, enforceable standards for environmental review and public disclosure, rather than leaving them to be negotiated case by case. Until then, data center governance in the United States is likely to remain a patchwork, shaped as much by courtrooms and city council meetings as by
legislatures.

\subsection{Acknowledgements}

The legal research supporting this report was enabled by a \textit{pro bono} legal hackathon hosted by the Data \& Society Research Institute under the supervision of Brian J. Chen in March 2026. The legal review during this hackathon was in part inspired by an early version of the "North Start Data Center Policy Toolkit" published by AI Now.\footnote{\href{https://datacenters.ainowinstitute.org/guides/}{AI Now - Guides - "North Start Data Center Policy Toolkit"}}

\clearpage
\hypertarget{appendix}{%
\section{Appendix}} \label{appendix}

On a state level, important environmental regulations vary. The Code of
Virginia includes laws and regulations related to air, land, renewable
energy and water. In terms of air, the laws cover stationary sources,
such as industrial facilities and regulations to make sure projects meet
federal requirements. Renewable energy law requires the Department of
Environmental Quality (DEQ) to develop regulations for permitting small
renewable energy projects. Finally, water regulations are promulgated by
the State Water Control Board and include permits, fees, groundwater
management and withdrawals. The code also recognizes the existence of
citizen regulatory boards, such as: State Air Pollution Control Board,
State Water Control Board and Virginia Waste Management
Board.\footnote{\url{https://www.deq.virginia.gov/news-info/shortcuts/laws-regulations}}

Section §56-235 of the Code of Virginia states the importance of any
rate, toll charge or schedule of any public utility to be just and
reasonable.\footnote{\url{https://law.lis.virginia.gov/vacode/title56/chapter10/section56-235.2/}}
It's important to consider that Virginia offers a data center retail
sales and use tax exemption (DCRSUT Exemption) on qualifying computer
equipment or enabling software purchased or leased for use in data
centers which meet minimum investment and job creation
requirements.\footnote{\url{https://www.vedp.org/incentive/data-center-retail-sales-use-tax-exemption}}

Another example comes from the state of Minnesota. 2025 Minnesota
statutes 114C-116I include numerous environmental protections such as:
environmental regulatory innovations, environmental response and
liability, pollution control agency, public water and sewer systems,
environmental rights, critical areas, etc.\footnote{\url{https://www.revisor.mn.gov/statutes/part/ENVIRONMENTAL+PROTECTION}}
Similar to Virginia, in Minnesota, Statute §297A.68 subdivision 42
provides a sales tax exemption for data centers on purchases of
information technology equipment. There are requirements to qualify
related to size, infrastructure and investment.\footnote{\url{https://www.revisor.mn.gov/statutes/cite/297a.68}}

\end{document}